\documentclass[aap,preprint]{imsart}

\RequirePackage[OT1]{fontenc}
\RequirePackage{amsfonts,amsthm,amsmath}
\RequirePackage[numbers]{natbib}
\RequirePackage[colorlinks,citecolor=blue,urlcolor=blue]{hyperref}
\usepackage{pdflscape} 
\usepackage{graphics}
\usepackage{epsfig}
\usepackage{multirow}
\usepackage{float}

\arxiv{arXiv:0000.0000}

\startlocaldefs
\numberwithin{equation}{section}
\theoremstyle{plain}

\endlocaldefs

\begin{document}

\begin{frontmatter}
\title{Lee-Carter method for forecasting mortality for Peruvian Population}	
\runtitle{Modeling and Forecasting Mortality}

\begin{aug}
\author{\fnms{J.} \snm{Cerda-Hern\'andez}
\ead[label=e1]{josecehe@gmail.com}}
\and
\author{\fnms{A.} \snm{Sikov}
\ead[label=e2]{anna.sikov@mail.huji.ac.il }}


\runauthor{J. Cerda-Hernandez}

\affiliation{National Engineering University}

\address{National Engineering University\\
Department of Engineering Economics\\
\printead{e1}
\phantom{E-mail:\  josecehe@gmail.com}
\printead{e2}
}

\end{aug}

\begin{abstract}
In this article, we have modeled mortality rates of Peruvian female and male 
populations during the period of 1950-2017 using the Lee-Carter (LC) model. 
The stochastic mortality model was introduced by Lee and Carter (1992) and has been used 
by many authors for fitting and forecasting the human mortality rates. The Singular Value Decomposition (SVD) approach is used for 
estimation of the parameters of the LC model. Utilizing the best fitted auto regressive integrated moving average (ARIMA) model  we 
forecast the values of the time dependent 
parameter of the LC model  for the next thirty years. The forecasted values of 
life expectancy at different age group with $95\%$ confidence intervals are also reported for the next thirty years. 
In this research we use the data, obtained from the Peruvian National Institute of 
Statistics (INEI).
\end{abstract}



\end{frontmatter}

\section{Introduction}
Mortality rate is an important variable in the fields of actuarial science, demography, national planning and social security 
administration. Mortality levels are generally regarded as indicators of a general welfare of a  population. 
Large changes in mortality rates in a relatively  short period of time may present a number of challenges to 
demographers and practitioners of actuarial science. For example, in the Peruavian case the 
death rate has reduced to a large extent during the last few decades. Specifically, according to the World Health 
Organization’s health statistics 2014, life expectancy at birth has increased by six years between 1990 and 2012 universally 
(77 years in 2012 as opposed to 71 years in 1990). This arises the need to 
develop methods for forecasting mortality rates and life expectancy. Prediction of future mortality rates are 
especially useful for life insurance companies and annuity providers, which use these predicted mortality rates in their pricing 
calculations. Clearly, systematic underestimation the longevity risk may eventually cause a financial 
collapse of these companies. For example, if mortality rates increase, the life insurers need to pay the death benefits earlier than expected. 
This implies that dramatic decline in mortality brings very serious financial exposures for insurers providing life contracts and life annuities.

Lee and Carter \cite{lc1992US} introduced the first mortality model with stochastic forecast.  The LC model is a two-factor  model which includes two age-specific parameters for every age group, and a time-varying  effect, such that a tendency of all age-specific central death rates have the same pattern of stochastic evolvement over time.
There have been several extensions of the basic Lee-Carter model by including different factors. Both, Maindonald,
and Smith \cite{Booth2002} considered the multi factor age-period extension of Lee-Carter, Renshaw and Haberman 
\cite{Renshaw2006} proposed a model with the cohort effect and \cite{Cairns2006} used the logit transformation in the 
mortality model. The main aim of this study is to fit the LC model for predicting Peruvian mortality 
rates and life expectancy in different age groups. We use life table data from 1950 to 2017. The central mortality rates were 
measured once during each 5 years period.  Based on the LC model, we predict central mortality rates and values of life expectancy 
at different age groups for the next six periods of five years, starting from the period of 2020-2025. 

The rest of the paper is organized as follows. In the next section we describe the data obtained from INEI
and give a brief discussion of the mortality pattern in Peru. In Section 3 we present the Lee Carter model and describe the 
estimation and forcasting procedure. In Section 4 we report the results of fitting the Lee Carter model to the Peruvian data.
In Section 5 we present the forecasting results. In Section 6 some conclusions are outlined.

\section{Data Description}

The  age-specific central mortality  rates  from 1950 to 2017 are available from the 
Peruvian National Institute of Statistics (INEI). There are 14 measurements for each age group: the first 
measurement refers to the period of 1950-1955, the second measurement refers to the period of 1955-1960, and so on. 
The last measurement is based on the census, which was conducted in Peru in November of 2017 (referred to as a period of 
2015-2020). The data are available for 18 age groups: 0, 1-4, 5-9, 10-14,...,75-79 and 80+. 
Unfortunately, such a layout of the data is insufficient for deriving some monetary functions involving life 
contingencies, since this generally requires knowledge of probabilities of death for every single year of age.
In the case of mortality at advanced ages the  INEI does not have detailed information; the only information
available is the central mortality rate at the age group of $80+$. 
There exist various mortality prediction models for the advanced ages (see for example \cite{Horiuchi1998}, \cite{Thatcher1999}),
however, making analysis  of the behavior of mortality rates at advanced ages is not the focus of this research.

The raw data, used for the purpose of implementation of the LC methodology are presented in Tables  \ref{mortalitymale}  and \ref{mortalityfemale}.  

\begin{figure}[h!] 
	\includegraphics[width=12cm]{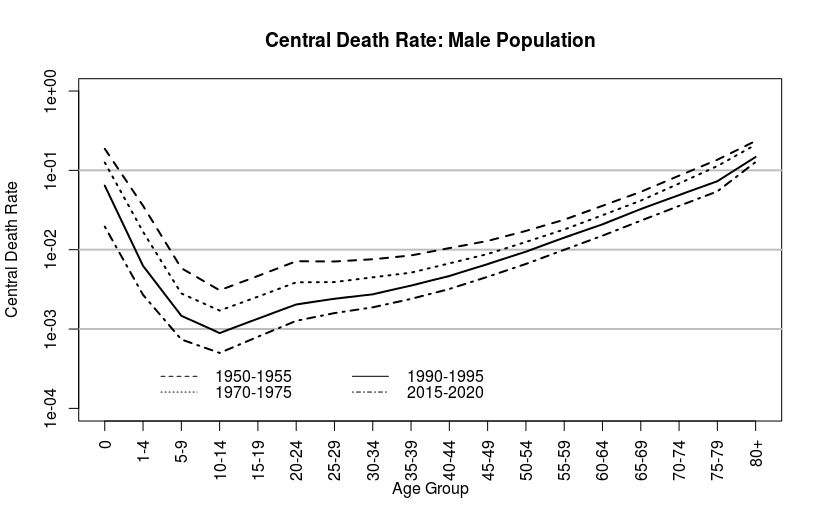}
	\caption{The central mortality rates for Peruvian male population.}
	\label{cdr_m}
\end{figure}

\begin{figure}[h!] 
	\includegraphics[width=12cm]{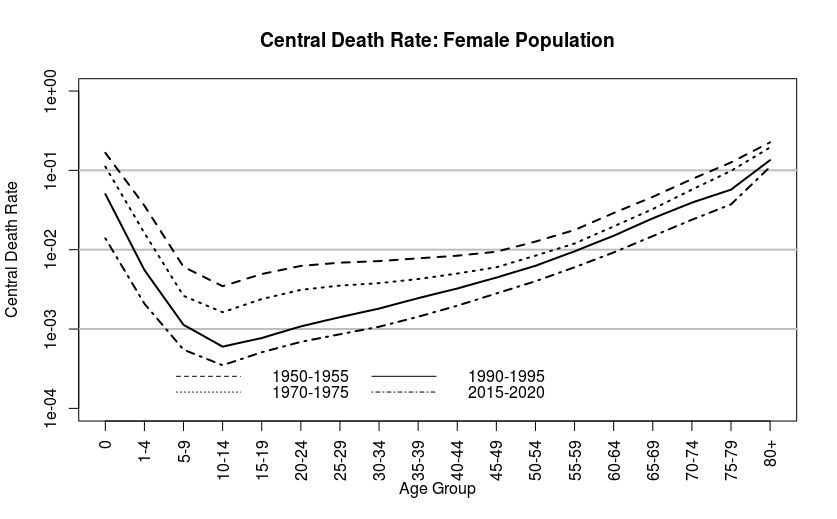}
	\caption{The central mortality rates for Peruvian female population.}
	\label{cdr_f}
\end{figure}

\medskip

Figures \ref{cdr_m} and \ref{cdr_f} present the age group specific central mortality rates for Peruvian female and 
male populations for 4 different periods: 1950-1955, 1970-1975, 1990-1995 and 2015-2020. These figures shows very 
clearly a notable reduction in mortality rates in Peru over time for both male and female populations. 
One can also observe that a more conciderable decline in mortality rates occurs in the younger ages groups. For the female 
population we observe a more rapid decline in mortality rates for the age groups from 10 to 40 years, during 1970 to 1995 as 
compared to the periods between 1950 to 1975 and between 1995-2015, while for the male population for the mentioned periods the
decline is uniform.
We can also conclude that during the periods between 1950 and 1975 and between 1975 and 1995, a more conciderable reduction 
occured for the age groups between 5 and 40, compared to the older age groups. During the last period between 1995 and 2015
the decline is generally more or less uniform for all the age groups.

\begin{figure}[h!] 
	\includegraphics[width=12cm]{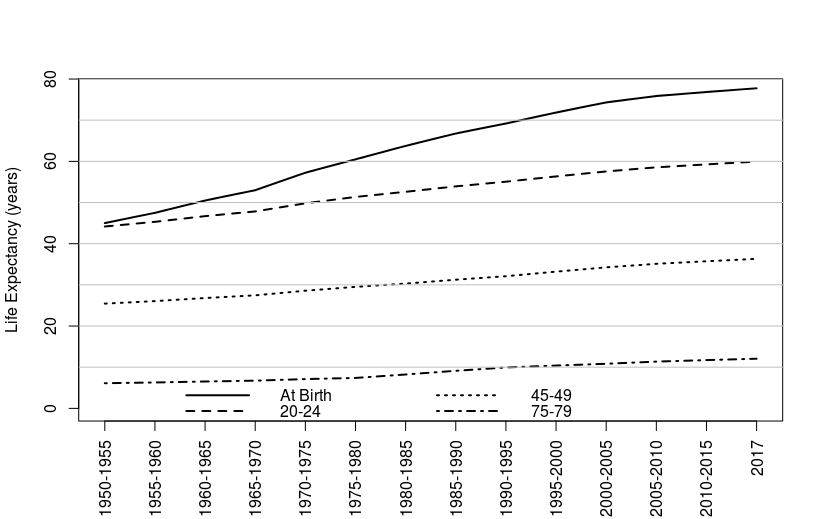}
	\caption{Life expectancy (in years) of Peruvian female population during 1950-2017 at selected age groups}
	\label{expectancy_f}
\end{figure}

\begin{figure}[h!] 
	\includegraphics[width=12cm]{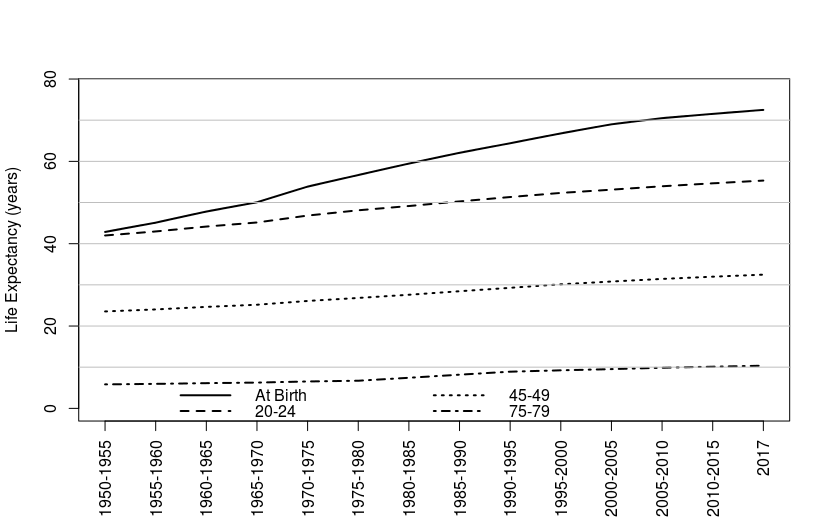}
	\caption{Life expectancy (in years) of Peruvian male population during 1950-2017 at selected age groups}
	\label{expectancy_m}
\end{figure}

Based on the data on mortality tables we compute life expectancies for female and 
male populations for all available 5 years periods, for several age groups. The results
are presented in Figures \ref{expectancy_f} and \ref{expectancy_m}. The figures illustrate an increase in life expectancy for all age groups for both
male and female populations: from drastic for the infants (72.5 and 77.8 years in 2017 as opposed to 42.9 and 45.0 in 1950-1955)
to quite modest for the 75-79 age group (10.4 and 12.1 in 2017 as opposed to 5.8 and 6.1 in 1950-1955).
One can also observe that in the period between 1950 and 1960, life expectancy of Peruvian female and male
at birth is quite close to life expectancy in the group of 20-24 .This can be explained by high rate of infant mortality 
during this decade.

\section{Lee-Carter Model}

Let $m_{x,t}$ denote the central mortality rate for the age group $x$, during the five years period $t$, where\\
$x\in \{0, 1-4, 5-9, 10-14,15-19,20-24, 25-29, 30-34, 35-39, 40-44, 45-49, 50-54, 55-59, 60-64, 65-69, 70-74, 75-79,80+\}$\\
and\\
$t\in \{1950-1955, 1955-1960,1960-1965, 1965-1970, 1970-1975, 1975-1980, 1980-1985, 1985-1990, 1990-1995, 
1995-2000, 2000-2005, 2005-2010, 2010-2015, 2015-2020\}$

The LC model uses the natural logarithm of the central mortality rates to measure the age and time 
effect, and is defined as

\begin{equation}\label{lc}
r_{x,t}=\ln ( m_{x,t}) = \alpha_x +  \beta_x k_t +  \varepsilon_{x,t}
\end{equation}
where  $\alpha_x$ denotes the coefficient which describes average age specific pattern by age of mortality, $k_t$ denotes the 
time-varying index for the general mortality,  $\beta_x$ denotes the coefficient which measures sensitivity of  
$\ln ( m_{x,t})$ at age group $x$ to changing the index $k_t$ (note that $d\ln(m_{x,t})/dt = \beta_xdk_t/dt$) and  
$\varepsilon_{x,t}$ is the error term which is assumed to follow a normal distribution 
with mean zero and to be independent of age and time. 
The  term  $\beta_xk_t$ in the LC model capture the joint tendency of age-specific mortality rates 
to evolve over time. 

The model can not be adjusted by regression methods since no explanation variables are included into the model. 
Moreover, the model is not identifiable (see Lee and Carter, 1992). 
In order to solve this problem, the authors use the following constraints: $\sum_t k_t=0$ and $\sum_x \beta_x=1$.  
The first constraint implies that $\alpha_x$  is equal to the average of  $\ln ( m_{x,t})$  over time. That is,
$$\hat{\alpha}_x = \frac{1}{T} \sum_{t=1}^{T}\ln ( m_{x,t})$$
where $T$ es the number of available time periods (in our case, $T=14$).  We therefore rewrite the model in terms of the mean 
centered log-mortality rate,  $\widetilde{r}_{x,t}=r_{x,t}-\overline{r}_{x,t}$. 
Since practical uses of the LC model implicitly assume that the disturbances $ \varepsilon_{x,t}$  are normally distributed, 
the Equation (\ref{lc}) can be expressed as a multiplicative fixed effects model for the centered age profile:
\begin{equation}\label{lc1}
\begin{array}{ccl}
\widetilde{r}_{x,t} & \sim & N\left( \hat{\alpha}_x, \sigma^2 \right) \\
\widetilde{r}_{x,t} & = & \beta_x k_t
\end{array}
\end{equation}
where the parameter $\hat{\alpha}_x=E(\widetilde{r}_{x,t})$  is interpreted as the average pattern of mortality at age $x$. Using constraints of the model, we obtain an estimate  of $k_t$, $k_t=\sum_x \ln ( m_{x,t}) -  \hat{\alpha}_x$. Differentiating both sides of (\ref{lc})  we obtain an estimate for $\beta_x$, $\hat{\beta}_x=(\partial \ln ( m_{x,t})/\partial t)/ (\partial k_t/\partial t )$.  

In order to estimate parameters of the LC model, Lee and Carter used Singular Values Decomposition (SVD) (see \cite{Lawson1974}, \cite{Shores2004})  of the matrix  $M_{x,t}= \ln ( m_{x,t}) -  \hat{\alpha}_x$ to obtain  $\beta_x$ and $k_t$:
\begin{equation}\label{sdv}
svd(M_{x,t}) =  \sum_{i=1}^{r} \lambda_i U_{x,i}V_{i,t}
\end{equation}
where $r=rank(M_{x,t})$, $\{\lambda_1 \geq  \lambda_2 \geq  \cdots \geq \lambda_r \}$ are the ordered singular values 
of $M_{x,t}$, $U_{x,i}$ and $V_{t,i}$ are the left and right singular vectors.  Utilizing the theorem of low rank 
approximation, the rank $h$ least square approximation of (\ref{sdv}) is obtained as
\begin{equation}
M_{x,t}^{(h)} =  \sum_{i=1}^{h} \lambda_i U_{x,i}V_{i,t} = \sum_{i=1}^{h}\beta_x^{(i)}k_t^{(i)}, \;\;\; h\leq r
\end{equation}
where  $\beta_x^{(i)}k_t^{(i)}= \lambda_i U_{x,i}V_{i,t} $ (for more detail see \cite{Koissi2008}, \cite{Lawson1974} and \cite{Shores2004}). Then,  the rank $h$ 
residuals associated with (\ref{sdv}) are  
$$\varepsilon_{x,t}=\sum_{i=h+1}^{r} \lambda_i U_{x,i}V_{i,t} =U\left[
\begin{array}{ccccccc}
0 & \cdots & 0 & \cdots & 0 & \cdots & 0 \\
\vdots & \ddots & \vdots & \ddots& \vdots & \ddots & \vdots \\
0 & \cdots & \lambda_{h+1} & \cdots & 0 & \cdots & 0 \\
\vdots & \ddots & \vdots & \ddots& \vdots & \ddots & \vdots \\
0 & \cdots & 0 & \cdots & \lambda_{r} & \cdots & 0 \\
\end{array}\right]V
$$ 
and the corresponding    rank-$h$ approximation least square errors is   $\varepsilon_{h}^2=\sum_{i=h+1}^{r} \lambda_i^2$ which 
implies that the errors have similar variance. However, this assumption is violated for mortality data: the variance of the 
log-central death rate is approximately  $Var(\ln(m_{x,t}))\approx 1/d_{x,t}$, where $d_{x,t}$ denote the 
number of deaths at the age group $x$ at time $t$ (see \cite{Wilmoth1993} for details). The proportion of variance explained by the $i^{th}$ term  $\lambda_i U_{x,i}V_{i,t}$ of the decomposition (\ref{sdv}) is given by  $\lambda_i^2/\sum_{j=1}^{r} \lambda_j^2$, and the total variance explained by a rank-$h$ approximation is  $\sigma_h^2= \sum_{i=1}^{h}\lambda_i^2/\sum_{j=1}^{r} \lambda_j^2$.  It
is clear that $0\leq \sigma_h^2 \leq 1 $ and   the closer this value is to 1, the better is the approximation.  
For example, for the US data, Lee and Carter \cite{lc1992US} restrained the SVD approximation to the first order 
$M_{x,t}^{(1)} \approx  \lambda_1 U_{x,1}V_{1,t}=\beta_x^{(1)}k_t^{(1)}$   
and obtained an explained variance $\sigma_1^2=92.7\%$ for the total population. 

Predicting mortality with the LC model is reduced to forecasting the index  $k_t$ utilizing time series approaches 
(see \cite{Hamilton1994}). 

\section{Fitting of the LC Model to Peruvian Mortality Data}
This section presents the results of estimation of parameters in LC model, described in the previous section,
for female and male populations of Peru, based on mortality tables available from the period of 1950-55 to the period of 2015-20. 
The estimated values of age dependent parameters $\alpha_x$ and $\beta_x$ are reported in Table \ref{ab} and the
estimated values of time dependent parameter $k_t$ are reported in Table \ref{kt}

\begin{figure}[h!] 
	\includegraphics[width=11.5cm]{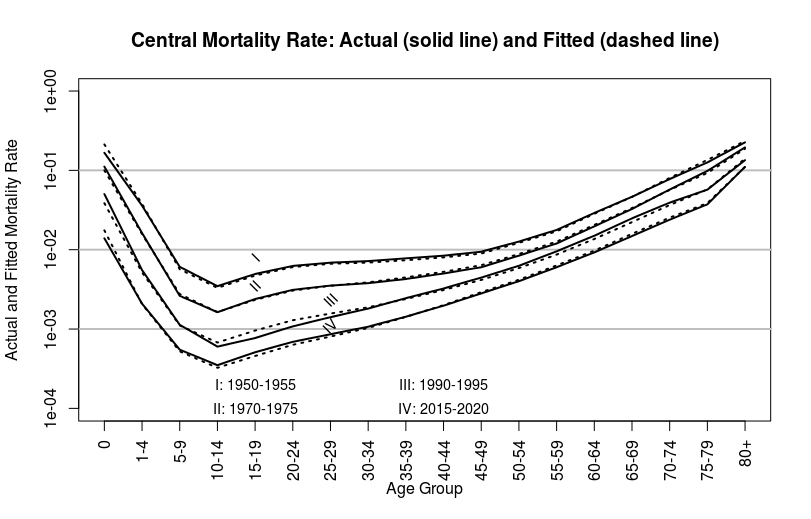}
	\caption{Actual and Fitted age group specific central mortality rates for Peruvian female population.}
	\label{fitted_f}
\end{figure}

\begin{figure}[H] 
	\includegraphics[width=11.5cm]{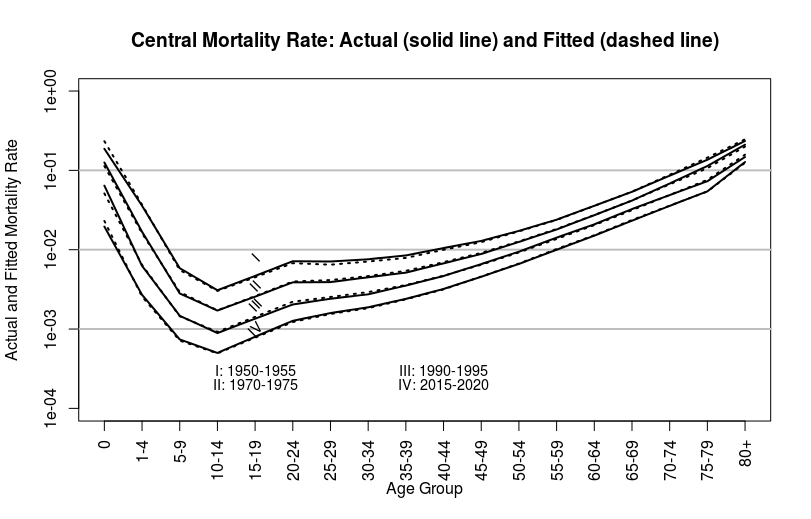}
	\caption{Actual and Fitted age group specific central mortality rates for Peruvian male population.} 
	\label{fitted_m}
\end{figure}

Applying SVD to the matrix $M_{x,t}$,  we  obtained an explained variance  of $98.73\%$  and $98.77\%$  by fitted LC model 
for Peruvian female and male mortality data respectively.  In Figures \ref{fitted_f} and \ref{fitted_m}, we have plotted 
the observed and fitted age group specific central mortality rates for four periods: 1950-1955, 1970-1975, 1990-1995 and 2015-2020. 
The obtained results indicate that the fitted mortality rates, obtained by fitting the LC 
model are generally very close to the observed (actual) mortality rates for both male and female 
populations although for the period of 1990-1995, the estimated mortality rates for females of the age 
groups 15-19 and 20-24 are somewhat higher than the actual mortality rates. Also there are some small differences for the 
newborns.


In Tables \ref{LEM} and \ref{LEF}, we present the actual values of LE and their estimated values, based on fitted 
LC for four selected decades.

\begin{table}[h!]
	\centering
	\caption{Estimates of  $\alpha_x$ and $\beta_x$ for Peruvian population based on quinquennial-wise mortality tables (1950-55 to 2015-20)}
		\label{ab}
	\begin{tabular}{|c|c|c|c|c|}
		\hline
		\multicolumn{1}{|c|}{\multirow{3}{*}{\begin{tabular}[c]{@{}c@{}}Age\\
		Group\vspace{0.4cm} \end{tabular}}} & \multicolumn{2}{c|}{Female} & \multicolumn{2}{c|}{Male} \\
		\cline{2-5}
		\multicolumn{1}{|c|}{}  & $\hat{\alpha}_x$ & $\hat{\beta}_x$ & $\hat{\alpha}_x$  & $\hat{\beta}_x$  
		\\ \hline
		0 & -2.8535 & 0.0825 & -2.6463 & 0.0946
		\\ \hline
		1-4 & -4.7967 & 0.0963 & -4.6714 & 0.1092
		\\ \hline
		5-9 & -6.4206 & 0.0789 & -6.2496 & 0.0838
		\\ \hline
		10-14 & -6.9271 & 0.0768 & -6.7344 & 0.0741
		\\ \hline
		15-19 & -6.5770 & 0.0772 & -6.3082 & 0.0717
		\\ \hline
		20-24 & -6.2901 & 0.0747 & -5.8756 & 0.0699
		\\ \hline
		25-29 & -6.1192 & 0.0699 & -5.7764 & 0.0585
		\\ \hline
		30-34 & -5.9718 & 0.0627 & -5.6453 & 0.0554
		\\ \hline
		35-39 & -5.7718 & 0.0545 & -5.4624 & 0.0492
		\\ \hline
		40-44 & -5.5531 & 0.0459 & -5.1991 & 0.0470
		\\ \hline
		45-49 & -5.3026 & 0.0371 & -4.9001 & 0.0408
		\\ \hline
		50-54 & -4.9664 & 0.0355 & -4.5534 & 0.0382
		\\ \hline
		55-59 & -4.5889 & 0.0327 & -4.1751 & 0.0345
		\\ \hline
		60-64 & -4.1227 & 0.0354 & -3.7713 & 0.0348
		\\ \hline
		65-69 & -3.6327 & 0.0355 & -3.3433 & 0.0336
		\\ \hline
		70-74 & -3.1249 & 0.0382 & -2.8910 & 0.0367
		\\ \hline
		75-79 & -2.6545 & 0.0414 & -2.4385 & 0.0403
		\\ \hline
		80+ & -1.8416 & 0.0249 & -1.7508 & 0.0278
		\\ \hline
	\end{tabular}
\end{table}

\begin{table}[h!]
	\centering
	\caption{$k_t$ for Peruvian population based on quinquennial wise mortality tables (1950-55 to 2015-20)}
	\label{kt}
	\begin{tabular}{|c|c|c|c|c|c|c|c|}
		\hline
		$t$ & 1950-55 & 1955-60 & 1960-65 & 1965-70 & 1970-75 & 1975-80& 1980-85\\
		\hline
		$\hat{k}_t$(female) & 15.826 & 14.176 & 12.082 & 10.216 & 6.697 & 3.648 & 0.777\\
		$k_t$(male) & 12.598 & 11.193 & 9.430 & 7.864 & 4.980 & 2.637 & 0.606\\
		\hline
	\end{tabular}
	
	\begin{tabular}{|c|c|c|c|c|c|c|c|}
		\hline
		$t$ & 1985-90 & 1990-95 & 1995-00& 2000-05 & 2005-10 & 2010-15 & 2015-20\\
		\hline
		$\hat{k}_t$(female) & -2.195 & -4.841 & -7.277 & -9.920 & -11.753 & -13.043 & -14.391\\
		$k_t$(male) &-1.448 & -3.419 & -5.463 & -7.473 & -9.148 & -10.504 & -11.853\\
		\hline
	\end{tabular}
\end{table}

\begin{table}[h!]
	\centering
	\caption{Observed and estimated life expectancy for the periods of 1950-1955, 1970-1975, 1990-1995 and 2015-2020, Males}
	\label{LEM}
	\begin{tabular}{|c|c|c|c|c|c|c|c|c|}
		\hline
		\multicolumn{1}{|c|}{\multirow{2}{*}{\begin{tabular}[c]{@{}c@{}}Age\\
		Group\end{tabular}}} & \multicolumn{2}{c|}{1950-1955} & \multicolumn{2}{c|}{1970-1975}  & \multicolumn{2}{c|}{1990-1995}  & \multicolumn{2}{c|}{2015-2020}\\
		\cline{2-9}
		\multicolumn{1}{|c|}{}  & Observed & Estimated & Observed  & Estimated  & Observed & Estimated  & Observed & Estimated 
		\\ \hline
		0 & 42.57 & 40.79 & 53.73 & 54.42 & 64.33 & 64.94 & 72.49 & 72.32
		\\ \hline
		1-4 & 50.24 & 50.39 & 59.88 & 59.90 & 67.57 & 67.33 & 72.91 & 73.00
		\\ \hline
		5-9 & 53.67 & 54.13 & 59.91 & 59.76 & 65.23 & 65.04 & 69.68 & 69.73
		\\ \hline
		10-19 & 50.19 & 50.58 & 55.72 & 55.61 & 60.70 & 60.49 & 64.93 & 64.97
		\\ \hline
		15-19 & 45.93 & 46.31 & 51.18 & 51.07 & 55.95 & 55.76 & 60.09 & 60.12
		\\ \hline
		20-24 & 41.96 & 42.31 & 46.81 & 46.70 & 51.32 & 51.14 & 55.32 & 55.35
		\\ \hline
		25-29 & 38.40 & 38.68 & 42.68 & 42.59 & 46.82 & 46.68 & 50.65 & 50.67
		\\ \hline
		30-34 & 34.70 & 34.87 & 38.47 & 38.43 & 42.35 & 42.25 & 46.04 & 46.05
		\\ \hline
		35-39 & 30.94 & 31.04 & 34.29 & 34.28 & 37.90 & 37.83 & 41.45 & 41.45
		\\ \hline
		40-44 & 27.17 & 27.19 & 30.11 & 30.15 & 33.54 & 33.47 & 36.92 & 36.91
		\\ \hline
		45-49 & 23.50 & 23.46 & 26.06 & 26.13 & 29.27 & 29.21 & 32.47 & 32.46
		\\ \hline
		50-54 & 19.90 & 19.80 & 22.12 & 22.24 & 25.16 & 25.09 & 28.17 & 28.16
		\\ \hline
		55-59 & 16.47 & 16.35 & 18.39 & 18.54 & 21.26 & 21.16 & 24.03 & 24.03
		\\ \hline
		60-64 & 13.25 & 13.10 & 14.89 & 15.07 & 17.64 & 17.48 & 20.13 & 20.16
		\\ \hline
		65-69 & 10.37 & 10.18 & 11.71 & 11.93 & 14.31 & 14.09 & 16.51 & 16.56
		\\ \hline
		70-74 & 7.83 & 7.60 & 8.86 & 9.14 & 11.41 & 11.08 & 13.24 & 13.34
		\\ \hline
		75-79 & 5.78 & 5.52 & 6.50 & 6.82 & 8.90 & 8.49 & 10.35 & 10.49
		\\ \hline
		80+ & 4.26 & 4.06 & 4.74 & 5.01 & 6.78 & 6.33 & 7.84 & 8.01
		\\ \hline
	\end{tabular}
\end{table}

\begin{table}[h!]
	\centering
	\caption{Observed and estimated life expectancy for the periods of 1950-1955, 1970-1975, 1990-1995 and 2015-2020, Females}
	\label{LEF}
	\begin{tabular}{|c|c|c|c|c|c|c|c|c|}
		\hline
		\multicolumn{1}{|c|}{\multirow{2}{*}{\begin{tabular}[c]{@{}c@{}}Age\\
		Group\end{tabular}}} & \multicolumn{2}{c|}{1950-1955} & \multicolumn{2}{c|}{1970-1975}  & \multicolumn{2}{c|}{1990-1995}  & \multicolumn{2}{c|}{2015-2020}\\
		\cline{2-9}
		\multicolumn{1}{|c|}{}  & Observed & Estimated & Observed  & Estimated  & Observed & Estimated & Observed & Estimated 
		\\ \hline
		0 & 44.76 & 42.75 & 57.13 & 57.75 & 69.15 & 70.25 & 77.75 & 77.22
		\\ \hline
		1-4 & 51.79 & 51.77 & 62.83 & 62.78 & 71.70 & 72.00 & 77.84 & 77.58
		\\ \hline
		5-9 & 55.50 & 55.93 & 62.91 & 62.73 & 69.26 & 69.47 & 74.48 & 74.21
		\\ \hline
		10-14 & 52.13 & 52.47 & 58.71 & 58.57 & 64.64 & 64.84 & 69.67 & 69.40
		\\ \hline
		15-19 & 48.00 & 48.30 & 54.17 & 54.03 & 59.82 & 60.05 & 64.79 & 64.51
		\\ \hline
		20-24 & 44.14 & 44.40 & 49.79 & 49.64 & 55.05 & 55.33 & 59.95 & 59.65
		\\ \hline
		25-29 & 40.46 & 40.68 & 45.53 & 45.36 & 50.33 & 50.67 & 55.15 & 54.83
		\\ \hline
		30-34 & 36.79 & 36.97 & 41.30 & 41.12 & 45.67 & 46.05 & 50.38 & 50.04
		\\ \hline
		35-39 & 33.05 & 33.18 & 37.04 & 36.88 & 41.06 & 41.46 & 45.63 & 45.29
		\\ \hline
		40-44 & 29.26 & 29.33 & 32.79 & 32.66 & 36.54 & 36.93 & 40.94 & 40.59
		\\ \hline
		45-49 & 25.41 & 25.43 & 28.56 & 28.46 & 32.09 & 32.47 & 36.32 & 35.98
		\\ \hline
		50-54 & 21.52 & 21.48 & 24.35 & 24.31 & 27.76 & 28.10 & 31.80 & 31.47
		\\ \hline
		55-59 & 17.76 & 17.68 & 20.29 & 20.29 & 23.56 & 23.86 & 27.39 & 27.08
		\\ \hline
		60-64 & 14.18 & 14.03 & 16.39 & 16.45 & 19.59 & 19.81 & 23.15 & 22.87
		\\ \hline
		65-69 & 11.01 & 10.80 & 12.83 & 12.97 & 15.93 & 16.03 & 19.13 & 18.89
		\\ \hline
		70-74 & 8.26 & 7.99 & 9.67 & 9.89 & 12.71 & 12.63 & 15.42 & 15.25
		\\ \hline
		75-79 & 6.07 & 5.79 & 7.09 & 7.36 & 9.93 & 9.68 & 12.06 & 11.98
		\\ \hline
		80+ & 4.43 & 4.25 & 5.13 & 5.34 & 7.43 & 7.11 & 9.03 & 9.02
		\\ \hline
	\end{tabular}
\end{table}

The results presented in Tables \ref{LEM} and \ref{LEF} show a very good fit of the LC model to the data for
both female and male populations, for all age groups.

\section{Forecasting}
Forecasting is generally the main aim behind the modeling of mortality rates. The notable advantage of the LC model is its
simplicity for predicting the future values of central mortality rates and life expectancy, since the values of the coefficients $a_x$ and $\beta_x$ are supposed to be constant over time.  
It follows then that in order to predict the future values of the mortality rate (and the life expectancy) one has to predict 
the corresponding value of the mortality index $k(t)$. In practice, for modeling the $k(t)$ the ARIMA models are
generally fitted. For example, Lee and Carter (1992) fitted ARIMA(0,1,0) (i.e. random walk with drift) for
modeling the mortality index for US population, Chavhan and Shinde (2016) utilized the ARIMA(1,2,0) and ARIMA(0,2,0) for
modeling the mortality index for female and male populations in India. Having fitted an appropriate model to the series 
of the observed values of $k(t)$, one can predict its future values, and consequently, compute predictions of the age specific central mortality 
rates and life expectancy, using the obtained values of $a_x$ and $b_x$ (see table \ref{ab}).
We considered a variety of ARIMA models to be fitted to mortality index for male and female populations. 
In both cases the best fitted model was ARIMA(0,2,0). The following tables present the results of estimation and forecasting
procedures. In Table \ref{kt_forecast}, we report the predicted values of 
mortality index, along with their corresponding standard errors for the next six periods (from 2020-2025 to 2045-2050). 
The results presented in the table show a steady reduction in predicted mortality rates over time for all age groups for both
males and females. Also, one can observe that a more rapid reduction occurs in the younger age groups. For example, for the
newborns, a reduction is about 43\% (from 1573 in 2020-2025 to 902 in 2045-2050) for females and about 47\% for males
(from 2034 in 2020-2025 to 1074 in 2045-2050); for the age group 45-49 a reduction is 22\% and 24\% for females and males 
correspondingly, and for the age group of 80+ a reduction is quite modest: 15\% and 17\%.

\begin{table}[h!]
	\centering
	\caption{Predicted values of mortality index $k_t$ with standard errors (in the parenteses), based on ARIMA (0,2,0) model
	for the next 6 quinnquenials (2020-2025 to 2045-2050)}
	\label{kt_forecast}
	\begin{tabular}{|c|c|c|c|c|c|c|}
		\hline
		$t$ & 2020-2025 & 2025-2030 & 2030-2035 & 2035-2040 & 2040-2045 & 2045-2050\\
		\hline
		$\hat{k}_t$(female) & -15.740(0.605) & -17.088(1.353) & -18.436(2.263) & -19.784(3.313) & -21.132(4.486) & -22.480(5.770) \\
		$k_t$(male) & -13.203(0.459) & -14.552(1.026) & -15.902(1.717) & -17.251(2.513) & -18.600(3.402) & -19.950(4.376) \\
		\hline
\end{tabular}
\end{table}

Table \ref{predmrates} presents the forecasted age specific central death rates in terms of deaths per 100,000 for the next 
six periods.

\begin{table}[h!]
	\centering
	\caption{Forecasted values of age specific mortality rates in terms per 100,000 for 2020-2025 (I),
	2025-2030 (II), 2030-2035 (III), 2035-2040 (IV), 2040-2045 (V) and 2045-2050 (VI)}
		\label{predmrates}
	\begin{tabular}{|c|c|c|c|c|c|c|c|c|c|c|c|c|}
		\hline
		\multicolumn{1}{|c|}{\multirow{3}{*}{\begin{tabular}[c]{@{}c@{}}Age\\
		Group\end{tabular}}} & \multicolumn{6}{c|}{Female} & \multicolumn{6}{c|}{Male} \\
		\cline{2-13}
		\multicolumn{1}{|c|}{}  & I & II & III  & IV  & V & VI
		& I & II & III  & IV  & V & VI
		\\ \hline
		0 & 1573 & 1407 & 1259 & 1127 & 1008 & 902 & 2034 & 1790 & 1575 & 1387 & 1220 & 1074
		\\ \hline
		1-4 & 181 & 159 & 140 & 123 & 108 & 95 & 221 & 191 & 165 & 142 & 123 & 106
		\\ \hline
		5-9 & 47 & 42 & 38 & 34 & 31 & 28 & 64 & 57 & 51 & 46 & 41 & 36 
		\\ \hline
		10-14 & 29 & 26 & 24 & 21 & 19 & 17 & 45 & 40 & 37 & 33 & 30 & 27 
		\\ \hline
		15-19 & 41 & 37 & 34 & 30 & 27 & 25 & 71 & 64 & 58 & 53 & 48 & 44
		\\ \hline
		20-24 & 57 & 52 & 47 & 42 & 38 & 35 & 112 & 102 & 92 & 84 & 77 & 70
		\\ \hline
		25-29 & 73 & 67 & 61 & 55 & 50 & 46 & 143 & 132 & 122 & 113 & 104 & 96
		\\ \hline
		30-34 & 95 & 87 & 80 & 74 & 68 & 62 & 170 & 158 & 147 & 136 & 126 & 117
		\\ \hline
		35-39 & 132 & 123 & 114 & 106 & 99 & 92 & 222 & 208 & 194 & 182 & 170 & 159
		\\ \hline
		40-44 & 188 & 177 & 166 & 156 & 147 & 138 & 297 & 279 & 261 & 245 & 230 & 216
		\\ \hline
		45-49 & 278 & 264 & 251 & 239 & 227 & 216 & 434 & 411 & 389 & 368 & 348 & 330
		\\ \hline
		50-54 & 399 & 380 & 362 & 346 & 329 & 314 & 636 & 604 & 574 & 545 & 518 & 492
		\\ \hline
		55-59 & 607 & 581 & 556 & 532 & 509 & 487 & 975 & 930 & 888 & 848 & 809 & 772
		\\ \hline
		60-64 & 928 & 885 & 844 & 804 & 767 & 731 & 1453 & 1387 & 1323 & 1262 & 1204 & 1149
		\\ \hline
		65-69 & 1513 & 1442 & 1375 & 1310 & 1249 & 1191 & 2266 & 2166 & 2070 & 1978 & 1890 & 1806
		\\ \hline
		70-74 & 2410 & 2289 & 2175 & 2066 & 1962 & 1864 & 3422 & 3257 & 3100 & 2950 & 2808 & 2672
		\\ \hline
		75-79 & 3664 & 3465 & 3276 & 3098 & 2930 & 2771 & 5131 & 4859 & 4602 & 4359 & 4129 & 3910
		\\ \hline
		80+ & 10717 & 10363 & 10021 & 9690 & 9371 & 9061 & 12028 & 11585 & 11159 & 10748 & 10352 & 9971
		\\ \hline
	\end{tabular}
\end{table}

Finally, Tables \ref{predle} and \ref{CIle} present the forecasted values of life expectancy and the corresponding 
95\% confidence intervals. From these tables one can observe that life expectancy at birth will increase from 77.75 to 81.98 for females and from 72.49 to 77.83 
for males (between the periods of 2015-2020 and 2045-2050). However, as one can see, the width of confidence intervals 
significantly increases for more distant periods. For example, for the period of 2045-2050 a confidence bands are around 4 years 
width for females and about 10 years width for males. For the period of 2020-2025 the band width is quite narrow: around 1 year
for females and around 5 years for males. For the younger age groups the confidence bands are generally wider. 

\begin{table}[h!]
	\centering
	\caption{Forecasted values of life expectancy for 2020-2025 (I),
	2025-2030 (II), 2030-2035 (III), 2035-2040 (IV), 2040-2045 (V) and 2045-2050 (VI)}
		\label{predle}
	\begin{tabular}{|c|c|c|c|c|c|c|c|c|c|c|c|c|}
		\hline
		\multicolumn{1}{|c|}{\multirow{3}{*}{\begin{tabular}[c]{@{}c@{}}Age\\
		Group \vspace{0.4cm} \end{tabular}}} & \multicolumn{6}{c|}{Female} & \multicolumn{6}{c|}{Male} \\
		\cline{2-13}
		\multicolumn{1}{|c|}{}  & I & II & III  & IV  & V & VI
		& I & II & III  & IV  & V & VI
		\\ \hline
		0 & 78.07 & 78.88 & 79.68 & 80.46 & 81.23 & 81.98 & 73.31 & 74.27 & 75.20 & 76.10 & 76.97 & 77.83
		\\ \hline
		1-4 & 78.30 & 79.00 & 79.69 & 80.37 & 81.04 & 81.72 & 73.81 & 74.60 & 75.39 & 76.15 & 76.91 & 77.67
		\\ \hline
		5-9 & 74.85 & 75.49 & 76.12 & 76.75 & 77.39 & 78.02 & 70.45 & 71.16 & 71.87 & 72.58 & 73.28 & 73.99
		\\ \hline
		10-14 & 70.02 & 70.64 & 71.26 & 71.88 & 72.50 & 73.12 & 65.67 & 66.36 & 67.05 & 67.74 & 68.43 & 69.12
		\\ \hline
		15-19 & 65.12 & 65.73 & 66.34 & 66.96 & 67.57 & 68.18 & 60.81 & 61.49 & 62.17 & 62.85 & 63.53 & 64.21
		\\ \hline
		20-24 & 60.25 & 60.85 & 61.45 & 62.05 & 62.66 & 63.26 & 56.01 & 56.68 & 57.34 & 58.01 & 58.67 & 59.34
		\\ \hline
		25-29 & 55.42 & 56.00 & 56.59 & 57.18 & 57.77 & 58.37 & 51.31 & 51.95 & 52.59 & 53.24 & 53.89 & 54.54
		\\ \hline
		30-34 & 50.61 & 51.18 & 51.75 & 52.33 & 52.91 & 53.50 & 46.66 & 47.28 & 47.90 & 48.53 & 49.16 & 49.79
		\\ \hline
		35-39 & 45.84 & 46.39 & 46.95 & 47.51 & 48.08 & 48.66 & 42.04 & 42.64 & 43.24 & 43.84 & 44.45 & 45.07
		\\ \hline
		40-44 & 41.13 & 41.66 & 42.21 & 42.75 & 43.31 & 43.87 & 37.48 & 38.05 & 38.63 & 39.22 & 39.81 & 40.41
		\\ \hline
		45-49 & 36.49 & 37.01 & 37.54 & 38.07 & 38.61 & 39.16 & 33.00 & 33.55 & 34.11 & 34.67 & 35.24 & 35.82
		\\ \hline
		50-54 & 31.97 & 32.47 & 32.98 & 33.50 & 34.02 & 34.55 & 28.67 & 29.20 & 29.73 & 30.27 & 30.82 & 31.38
		\\ \hline
		55-59 & 27.56 & 28.05 & 28.54 & 29.04 & 29.55 & 30.06 & 24.52 & 25.02 & 25.52 & 26.04 & 26.56 & 27.09
		\\ \hline
		60-64 & 23.33 & 23.80 & 24.27 & 24.75 & 25.24 & 25.74 & 20.62 & 21.09 & 21.57 & 22.06 & 22.55 & 23.06
		\\ \hline
		65-69 & 19.32 & 19.76 & 20.21 & 20.67 & 21.13 & 21.61 & 16.99 & 17.43 & 17.87 & 18.33 & 18.80 & 19.28
		\\ \hline
		70-74 & 15.65 & 16.06 & 16.47 & 16.90 & 17.33 & 17.78 & 13.73 & 14.14 & 14.55 & 14.98 & 15.42 & 15.87
		\\ \hline
		75-79 & 12.34 & 12.71 & 13.08 & 13.47 & 13.87 & 14.28 & 10.84 & 11.21 & 11.59 & 11.97 & 12.38 & 12.79
		\\ \hline
		80+ & 9.33 & 9.65 & 9.98 & 10.32 & 10.67 & 11.04 & 8.31 & 8.63 & 8.96 & 9.30 & 9.66 & 10.03 
		\\ \hline
	\end{tabular}
\end{table}

\begin{table}[h!]
	\centering
	\caption{Confidence Intervals for life expectancy for 2020-2025 (I),
	2030-2035 (III), and 2045-2050 (VI)}
		\label{CIle}
	\begin{tabular}{|c|c|c|c|c|c|c|}
		\hline
		\multicolumn{1}{|c|}{\multirow{3}{*}{\begin{tabular}[c]{@{}c@{}}Age\\
		Group \vspace{0.4cm} \end{tabular}}} & \multicolumn{3}{c|}{Female} & \multicolumn{3}{c|}{Male} \\
		\cline{2-7}
		\multicolumn{1}{|c|}{}  & I & III & VI
		& I & III & VI
		\\ \hline
		0 & 77.33 - 78.79 & 76.97 - 82.19 & 75.09 - 88.05 & 72.65 - 73.96 & 72.83 - 77.40 & 71.95 - 83.01
		\\ \hline
		1-4 &  77.67 - 78.91 & 77.37 - 81.91 & 75.82 - 87.36 & 73.27 - 74.34 & 73.41 - 77.29 & 72.70 - 82.40 
		\\ \hline
		5-9 & 74.29 - 75.41 & 74.02 - 78.2 & 72.66 - 83.47 & 69.97 - 70.92 & 70.09 - 73.63 & 69.47 - 78.53
		\\ \hline
		10-14 & 69.47 - 70.57 & 69.22 - 73.3 & 67.90 - 78.52 & 65.20 - 66.13 & 65.32 - 68.77 & 64.72 - 73.60
		\\ \hline
		15-19 & 64.58 - 65.66 & 64.33 - 68.36 & 63.03 - 73.55 & 60.35 - 61.26 & 60.47 - 63.86 & 59.88 - 68.65
		\\ \hline
		20-24 & 59.72 - 60.78 & 59.47 - 63.44 & 58.21 - 68.58 & 55.57 - 56.46 & 55.69 - 59.00 & 55.11 - 63.73
		\\ \hline
		25-29 & 54.90 - 55.93 & 54.66 - 58.54 & 53.44 - 63.63 & 50.89 - 51.74 & 51.00 - 54.21 & 50.45 - 58.85
		\\ \hline
		30-34 & 50.11 - 51.11 & 49.88 - 53.67 & 48.69 - 58.70 & 46.25 - 47.07 & 46.36 - 49.47 & 45.83 - 54.01
		\\ \hline
		35-39 & 45.35 - 46.33 & 45.13 - 48.82 & 43.99 - 53.78 & 41.65 - 42.44 & 41.75 - 44.76 & 41.24 - 49.20
		\\ \hline
		40-44 & 40.66 - 41.60 & 40.44 - 44.03 & 39.34 - 48.91 & 37.10 - 37.86 & 37.20 - 40.11 & 36.71 - 44.44
		\\ \hline
		45-49 & 36.04 - 36.95 & 35.83 - 39.32 & 34.77 - 44.10 & 32.64 - 33.37 & 32.74 - 35.53 & 32.27 - 39.75
		\\ \hline
		50-54 & 31.53 - 32.41 & 31.33 - 34.71 & 30.30 - 39.40 & 28.33 - 29.02 & 28.42 - 31.09 & 27.98 - 35.18
		\\ \hline
		55-59 & 27.14 - 27.99 & 26.94 - 30.21 & 25.96 - 34.79 & 24.19 - 24.85 & 24.28 - 26.82 & 23.86 - 30.77
		\\ \hline
		60-64 & 22.93 - 23.74 & 22.74 - 25.89 & 21.81 - 30.33 & 20.31 - 20.93 & 20.39 - 22.80 & 20.00 - 26.59
		\\ \hline
		65-69 & 18.94 - 19.71 & 18.77 - 21.75 & 17.89 - 26.02 & 16.7 - 17.28 & 16.78 - 19.04 & 16.41 - 22.64
		\\ \hline
		70-74 & 15.30 - 16.01 & 15.13 - 17.91 & 14.33 - 21.98 & 13.47 - 14.00 & 13.54 - 15.64 & 13.20 - 19.05
		\\ \hline
		75-79 & 12.02 - 12.66 & 11.88 - 14.40 & 11.16 - 18.20 & 10.61 - 11.09 & 10.67 - 12.58 & 10.36 - 15.75
		\\ \hline
		80+ & 9.06 - 9.61 & 8.94 - 11.14 & 8.33 - 14.62 & 8.11 - 8.52 & 8.16 - 9.84 & 7.90 - 12.73
		\\ \hline
	\end{tabular}
\end{table}

\begin{landscape}
\begin{table*}
	\caption{Quinquennial-wise age group specific central death rates for Peruvian male population during 1950-55 to 2015-20}
	\label{mortalitymale}
\tiny{
	\begin{tabular}{|c|c|c|c|c|c|c|c|c|c|c|c|c|c|c|}
		\hline
		
		Edad & 1950-55 & 1955-60 & 1960-65 & 1965-70 & 1970-75 & 1975-80& 1980-85 & 1985-90&1990-95 & 1995-00& 2000-05 & 2005-10 & 2010-15 & 2015-20\\
		\hline
		0 & 0.16642 & 0.15559& 0.14297& 0.13276& 0.11610& 0.10459& 0.08789& 0.07473& 0.06175& 0.04568& 0.03114& 0.02409& 0.02152& 0.01929\\
		\hline
		
		1-4 & 0.13037& 0.11479& 0.09773& 0.08460& 0.06437& 0.05111& 0.04001& 0.03117& 0.02448& 0.01968& 0.01533& 0.01357& 0.01208& 0.01074\\
		\hline
		5-9  & 0.02883& 0.02534& 0.02151& 0.01854& 0.01395& 0.01093& 0.00956& 0.00834& 0.00731& 0.00616& 0.00514& 0.00458& 0.00412& 0.00370\\
		\hline
		10-14 & 0.01528& 0.01375& 0.01202 & 0.01066& 0.00851& 0.00707& 0.00606& 0.00518&  0.00445& 0.00393& 0.00343& 0.00308& 0.00277& 0.00250 \\
		\hline
		15-19 & 0.02309 & 0.02073& 0.01808& 0.01600& 0.01274& 0.01056& 0.00907& 0.00777& 0.00669& 0.00605& 0.00547& 0.00490& 0.00443& 0.00399 \\
		\hline
		
		20-24 & 0.03519 & 0.03151& 0.02741& 0.02422& 0.01923& 06.01593& 0.01371& 0.01176& 0.01014& 0.00936 & 0.00866&  0.00777 & 0.00702& 0.00633\\
		\hline
		25-29 & 0.03492& 0.03137 & 0.02739 & 0.02425 & 0.01934&  0.01606& 0.01455& 0.01316& 0.01194& 0.01126& 0.01066& 0.00959& 0.00871& 0.00792\\ 
		\hline
		30-34 & 0.03720& 0.03383& 0.03001& 0.02700& 0.02222& 0.01902& 0.01701& 0.01517 & 0.01359& 0.01289 & 0.01226& 0.01115& 0.01020 & 0.00935\\
		\hline
		35-39 & 0.04146 & 0.03787 & 0.03379& 0.03055& 0.02539& 0.02191& 0.02032 & 0.01881& 0.01745& 0.01629& 0.01524& 0.01396& 0.01290& 0.01193 \\
		
		\hline
		40-44 & 0.05108 & 0.04712 & 0.04258& 0.03894& 0.03310& 0.02912 & 0.02692& 0.02485& 0.02300& 0.02129& 0.01974 & 0.01825& 0.01698& 0.01585\\
		\hline
		45-49 & 0.06255& 0.05836& 0.05350& 0.04959& 0.04321& 0.03882& 0.03652& 0.03431& 0.03229& 0.02983& 0.02760& 0.02566& 0.02404& 0.02257 \\
		
		\hline
		50-54 & 0.08279&  0.07810 & 0.07260& 0.06812& 0.06076& 0.05564 & 0.05223& 0.04896& 0.04600& 0.04134& 0.03914& 0.03661& 0.03447& 0.03255 \\
		\hline
		55-59 & 0.11270& 0.10709 & 0.10046& 0.09504& 0.08609& 0.07982& 0.07581& 0.07192& 0.06835& 0.06291 & 0.05799& 0.05441& 0.05140& 0.04869\\
		\hline
		60-64 & 0.16423& 0.15650 & 0.14737& 0.13988 & 0.12748& 0.11878& 0.11122& 0.10483& 0.09898& 0.09176 & 0.08523 & 0.08030& 0.07616& 0.07241\\
		\hline
		65-69 & 0.23669& 0.22658& 0.21459& 0.20474 & 0.18837& 0.17685& 0.16742& 0.15832& 0.14999& 0.13863& 0.12837& 0.12123& 0.11523& 0.10981\\
		\hline
		70-74 & 0.35286& 0.34018& 0.32507& 0.31260& 0.29178& 0.27704& 0.25541 & 0.23509& 0.21697& 0.20243& 0.18926& 0.17944 & 0.17115& 0.16367 \\
		
		\hline
		75-79 & 0.50799& 0.49446& 0.47823& 0.46475& 0.44207& 0.42589& 0.38235& 0.34253& 0.30797& 0.28924& 0.27232& 0.25962& 0.24892& 0.23926 \\
		\hline
		80+ & 1.00000& 1.00000& 1.00000& 1.00000& 1.00000& 1.00000& 1.00000& 1.00000& 1.00000& 1.00000& 1.00000& 1.00000& 1.00000& 1.00000\\
		\hline
	\end{tabular}
}
\end{table*}

\begin{table*}
		\caption{Quinquennial-wise age group specific central death rates for Peruvian female population during 1950-55 to 2015-20}
		\label{mortalityfemale}
\tiny{
	\begin{tabular}{|c|c|c|c|c|c|c|c|c|c|c|c|c|c|c|}
		\hline
		
		Edad & 1950-55 & 1955-60 & 1960-65 & 1965-70 & 1970-75 & 1975-80& 1980-85 & 1985-90&1990-95 & 1995-00& 2000-05 & 2005-10 & 2010-15 & 2015-20\\
		\hline
		0 &0.15044 &0.14047 &0.12886 &0.11947 &0.10416 &0.09327 &0.07507 &0.06096 &0.04883 &0.03531 &0.02346 &0.01775 &0.01558 &0.01379\\
		\hline
		1-4 &0.13078 &0.11464 &0.09693 &0.08327 &0.06215 &0.04786 &0.03667 &0.02787 &0.02174 &0.01620 &0.01135 &0.01008 &0.00917 &0.00832\\
		\hline
		5-9 &0.02996 &0.02593 &0.02155 &0.01819 &0.01303 &0.00955 &0.00798 &0.00663 &0.00560 &0.00458 &0.00369 &0.00328 &0.00300 &0.00274\\
		\hline
		10-14 &0.01720 &0.01507 &0.01273 &0.01093 &0.00814 &0.00625 &0.00487 &0.00376 &0.00298 &0.00265 &0.00236 &0.00210 &0.00192 &0.00175\\
		\hline
		15-19 &0.02441 &0.02145 &0.01820 &0.01571 &0.01187 &0.00927 &0.00689 &0.00508 &0.00385 &0.00361 &0.00340 &0.00304 &0.00279 &0.00255\\
		\hline
		20-24 &0.03079 &0.02721 &0.02326 &0.02021 &0.01547 &0.01225 &0.00929 &0.00698 &0.00539 &0.00496 &0.00459 &0.00412 &0.00377 &0.00346\\
		\hline
		25-29 &0.03378 &0.02999 &0.02580 &0.02254 &0.01748 &0.01402 &0.01109 &0.00871 &0.00700 &0.00626 &0.00561 &0.00504 &0.00465 &0.00427\\
		\hline
		30-34 &0.03533 &0.03154 &0.02730 &0.02399 &0.01878 &0.01521 &0.01273 &0.01060 &0.00897 &0.00789 &0.00694 &0.00627 &0.00578 &0.00533\\
		\hline
		35-39 &0.03822 &0.03434 &0.02998 &0.02655 &0.02113 &0.01740 &0.01540 &0.01358 &0.01213 &0.01055 &0.00918 &0.00832 &0.00771 &0.00714\\
		\hline
		40-44 &0.04120 &0.03754 &0.03337 &0.03005 &0.02474 &0.02103 &0.01919 &0.01746 &0.01603 &0.01416 &0.01250 &0.01137 &0.01056 &0.00981\\
		\hline
		45-49 &0.04606 &0.04246 &0.03831 &0.03498 &0.02960 &0.02582 &0.02446 &0.02312 &0.02198 &0.01966 &0.01762 &0.01607 &0.01498 &0.01394\\
		\hline
		50-54 &0.06142 &0.05706 &0.05200 &0.04791 &0.04125 &0.03652 &0.03448 &0.03249 &0.03079 &0.02763 &0.02486 &0.02274 &0.02126 &0.01985\\
		\hline
		55-59 &0.08484 &0.07910 &0.07240 &0.06699 &0.05815 &0.05187 &0.04997 &0.04809 &0.04645 &0.04134 &0.03686 &0.03381 &0.03163 &0.02960\\
		\hline
		60-64 &0.13536 &0.12642 &0.11600 &0.10756 &0.09374 &0.08389 &0.07970 &0.07561 &0.07208 &0.06350 &0.05597 &0.05140 &0.04817 &0.04514\\
		\hline
		65-69 &0.20791 &0.19572 &0.18142 &0.16978 &0.15063 &0.13691 &0.12971 &0.12268 &0.11665 &0.10174 &0.08869 &0.08150 &0.07641 &0.07161\\
		\hline
		70-74 &0.32508 &0.30937 &0.29082 &0.27560 &0.25039 &0.23218 &0.21228 &0.19356 &0.17801 &0.15716 &0.13887 &0.12784 &0.12001 &0.11267\\
		\hline
		75-79 &0.47884 &0.46148 &0.44077 &0.42366 &0.39504 &0.37417 &0.32632 &0.28341 &0.24940 &0.22689 &0.20717 &0.19169 &0.18074 &0.17043\\
		\hline
		80+ &1.00000 &1.00000 &1.00000 &1.00000 &1.00000 &1.00000 &1.00000 &1.00000 &1.00000 &1.00000 &1.00000 &1.00000 &1.00000 &1.00000\\
		\hline
		
	\end{tabular}
}
\end{table*}
\end{landscape}

\section{Conclusions}

In this paper we illustrate the performance of the LC approach to modeling the central mortality rates of Peruvian population. 
The principal objective of this study is to estimate the model parameters and predict future values
of central mortality rates as well as future life expectancy. The data for central mortality rates is available for 14 five 
years periods (census data), from 1950 to 2017. As mentioned above, these predictions are utilized by life insurance companies
and annuity providers for their pricing calculations. 
The results, presented in this article demonstrate a very good fit of the model to the data. On the other hand, the confidence
intervals for life expectancy, presented in Table \ref{CIle} are somewhat wide for more distant periods, especially for
the male population. This can probably be explained by a large variability of the mortality index predictions due to a shortness of the
series of the mortality index (recall that in our case it is only 14). Since the insurance company are interested in long 
term predictions, the width of confidence intervals can be of great importance. The authors are working in this direction.


\end{document}